\documentclass{article}
\usepackage{spconf,amsmath,graphicx}
\usepackage{multirow, tabularx}
\usepackage{booktabs}
\usepackage{xcolor}
\usepackage{amssymb}
\usepackage{enumitem}
\setitemize{noitemsep,topsep=0pt,parsep=0pt,partopsep=0pt}
\usepackage{algorithm}
\usepackage{fnpos}
\usepackage[colorlinks=false]{hyperref}

\usepackage{algpseudocode}

\title{Advancing Acoustic Howling Suppression through Recursive Training of Neural Networks}

\name{Hao Zhang$^{1*}$, Yixuan Zhang$^{2*}$, Meng Yu$^1$, Dong Yu$^1$\thanks{$^*$Equal contributions by H. Zhang and Y. Zhang. This work was performed when Y. Zhang was an intern at Tencent AI Lab.}}
\address{$^1$Tencent AI Lab, Bellevue, WA, USA\\
$^2$The Ohio State University, Columbus, OH, USA}
\begin{document}
\ninept
\maketitle
\begin{abstract}
In this paper, we introduce a novel training framework designed to comprehensively address the acoustic howling issue by examining its fundamental formation process.
This framework integrates a neural network (NN) module into the closed-loop system during training with signals generated recursively on the fly to closely mimic the streaming process of acoustic howling suppression (AHS).
The proposed recursive training strategy bridges the gap between training and real-world inference scenarios, marking a departure from previous NN-based methods that typically approach AHS as either noise suppression or acoustic echo cancellation.
Within this framework, we explore two methodologies: one exclusively relying on NN and the other combining NN with the traditional Kalman filter. Additionally, we propose strategies, including howling detection and initialization using pre-trained offline models, to bolster trainability and expedite the training process. 
Experimental results validate that this framework offers a substantial improvement over previous methodologies for acoustic howling suppression.

%

\end{abstract}
\section{Introduction}
\label{sec:intro}

Acoustic howling is a phenomenon stems from positive feedback within the audio system itself, often caused by the amplified sound output from the loudspeaker being picked up by the microphone and subsequently re-amplified \cite{waterhouse1965theory, van2010fifty, loetwassana2007adaptive}. 
This results in an uncontrolled positive feedback loop, leading to the undesirable amplification of specific frequency components and the generation of a sustained and unpleasant howling sound. It is commonly observed in systems like hearing aids, public addressing system, and karaoke. 
The presence of howling not only poses a threat to the functionality of the equipment but also poses potential risks to human hearing system.

Many methods have been proposed for acoustic howling suppression (AHS), including gain control \cite{schroeder1964improvement}, frequency shift \cite{berdahl2010frequency}, notch filter \cite{gil2009regularized, waterschoot2010comparative}, and adaptive feedback cancellation (AFC) \cite{joson1993adaptive, albu2018hybrid, waterschoot2010comparative}. Among them, the AFC method employs adaptive filters like the Kalman filter \cite{enzner2006frequency} to estimate and cancel howling signals by continuously updating filter coefficients based on detected feedback, making them a powerful approach for AHS over other methods. However, AFC techniques are sensitive to control parameters and often fall short in feedback systems exhibiting nonlinear distortions.

Acoustic howling is similar to acoustic echo since they both arise from feedback in communication systems and mishandling acoustic echo can lead to howling. 
Deep learning has demonstrated impressive performance in tackling acoustic echo problems \cite{zhang2018deep,  zhang2022neural} and has recently emerged as a promising solution for AHS tasks.
Chen et al. \cite{chen2022neural} introduced a deep learning method for howling detection. 
Later, two deep learning-based approaches for AHS were introduced: howling noise suppression \cite{gan2022howling} and deep marginal feedback cancellation (DeepMFC) \cite{zheng2022deep}. These methods treat AHS as a noise suppression task, training neural network (NN) modules to estimate target speech directly from offline generated microphone signals without incorporating AHS processing.
A distinct approach, known as DeepAHS \cite{zhang2023deepAHS}, leverages teacher-forcing learning and displays superior performance compared to previous methods. Building upon DeepAHS, HybridAHS \cite{zhang2023hybridAHS} further enhances howling suppression by incorporating the output of a Kalman filter as an additional input for model training.
Despite these strides, current methods all adhere to offline-generated signal training, leading to a mismatch between training and real-time inference that ultimately curtails their effectiveness.

\begin{figure}[!t]
\centering
     \includegraphics[width=0.95\columnwidth]{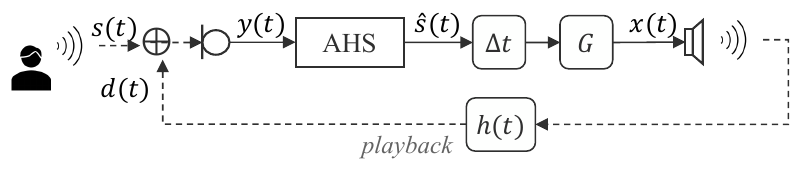}
      \caption{Configuration of an AHS system.}
      \label{fig:howling}
\end{figure}

This paper introduces an innovative training approach for acoustic howling suppression by implementing recursive training of a neural network. Specifically, we utilize a recurrent neural network with long short-term memory (LSTM) architecture \cite{graves2012long} as the NN module and integrate it into the closed-loop system of the howling suppression setup for frame-by-frame processing of the microphone signal. To achieve howling suppression, the NN module is trained to estimate the target speech from the microphone signal using complex ratio mask (cRM) estimation \cite{williamson2015complex}. Input signals for model training are generated recursively on the fly to emulate the fundamental process of acoustic howling formation, i.e., the output of the NN at each frame is passed through the closed-loop and subsequently fed back to generate its following input frames. 
To fully leverage the benefits of recursive training, we utilize either past processed loudspeaker signal or the output of Kalman filter as reference signals for training the NN module.

Our study offers three main contributions. Firstly, recursive training of the neural network effectively eliminates the mismatch limitations observed in previous NN-based AHS methods, leading to enhanced performance and increased robustness. Secondly, we explore two configurations within this framework: a pure NN-based method and a hybrid approach that combines NN with a Kalman filter, providing flexibility for the design of NN-based AHS systems. Thirdly, we employ strategies such as howling detection and initialization using pre-trained models to facilitate convergence of recursive training.

The remainder of this paper is organized as follows. Section 2 introduces the acoustic howling problem. Section 3 presents the proposed method. The experimental setup and results are described in Section 4 and 5, respectively. Section 6 concludes the paper.

\section{Acoustic Howling Suppression}

\subsection{Problem formulation of acoustic howling}

Let us consider a single-channel acoustic amplification system as shown in Fig.~\ref{fig:howling}. The target signal $s(t)$ captured by microphone is transmitted to the loudspeaker for acoustic amplification. The amplified signal $x(t)$ is played out by the loudspeaker and arrives at the microphone as playback $d(t)$, the corresponding microphone signal is a mixture of the target speech and the playback signal:
\begin{flalign}
\label{equ:1}
y(t) = s(t) +  x(t)*h(t)
\end{flalign}
where $*$ denotes linear convolution, $h(t)$ represents the acoustic path from loudspeaker to microphone. 


Without any AHS processing, the loudspeaker signal $x(t)$ will be an amplified version of the previous microphone signal $y(t-\Delta t)$ and undergo repeated re-entry into the pickup, leading to the representation of the microphone signal as:
\begin{flalign}
\label{equ:howling1}
\textstyle y(t) =   s(t) +   \left[y(t-\Delta t) \cdot G\right]*h(t)
\end{flalign}
where $\Delta t$ indicates the system delay from the microphone to the loudspeaker, and $G$ denotes the amplifier gain.
With proper howling suppression, the AHS module will suppress feedback and output an estimate of the target signal, $\hat{s}(t)$, and the corresponding microphone signal will be:
\begin{flalign}
\label{equ:howling2}
\textstyle y(t) =   s(t) +\left[\hat{s}(t-\Delta t) \cdot G\right]*h(t)
\end{flalign}
The recursive relationship between $y(t)$ and $y(t-\Delta t)$ and the possible leakage in $\hat{s}(t-\Delta t)$ give rise to the re-amplification of the playback signal, creating a feedback loop that manifests as an unpleasant, high-pitched sound known as acoustic howling.

Noted that the expression of AHS in equation (\ref{equ:1}) looks similar to the formulation of acoustic echo cancellation (AEC). 
%
But they differ in their origin and characteristics. Howling originates from the same source as the target signal and is generated gradually, making its suppression more challenging compared to acoustic echo, which is typically generated from a different source (far-end speaker) and can be treated as an instantaneous speech separation problem. 


\subsection{The mismatch problem in existing NN-based AHS methods}


The recursive nature of acoustic howling presents challenges in creating suitable training signals due to dependence between current inputs and previous outputs. Earlier NN-based techniques address AHS by training models with offline-generated microphone signals. Specifically, methods like howling noise suppression and DeepMFC utilize microphone signals generated offline without AHS, shown in equation (\ref{equ:howling1}). DeepAHS and HybridAHS adopt teacher forcing learning, assuming perfect howling suppression, where the microphone signals used for training are generated offline by substituting $\hat{s}(t)$ with $s(t)$ in equation (\ref{equ:howling2}).

The mismatch problem arises because the real microphone signal received during inference is generated recursively using the processed microphone signal, as described by equation (\ref{equ:howling2}), which differs from the training signals utilized in these methods.

\section{Proposed Method}

\subsection{Overall system}

\begin{figure}[!t]
\centering
     \includegraphics[width=0.85\columnwidth]{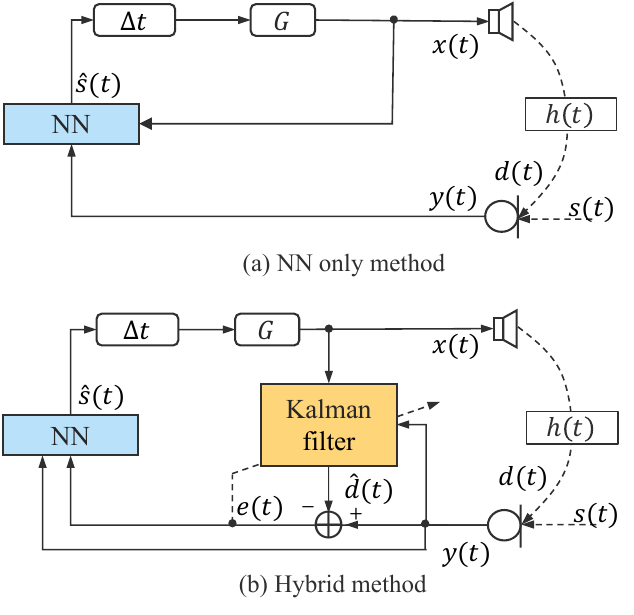}
      \caption{Diagrams of the proposed method (a) NN only method, (b) Hybrid method.}
      \label{fig:OnlineAHS}
\end{figure}


The recursive training approach is introduced to mitigate the mismatch problem and maintain consistency throughout training and inference. Here, input signals are generated recursively on the fly during model training where each processed frame serves as the input for the subsequent frame, preserving the inherent recursive nature of howling suppression.
Motivated by prior studies, we investigate configurations within our proposed approach that employ solely NN and hybrid NN with a Kalman filter for AHS, as shown in Fig.\ref{fig:OnlineAHS}.  The proposed method is implemented in the frequency domain while we employ time-domain labels in this figure to clarify signal relationships and improve comprehension. 
Further details of the proposed method are provided in Algorithm 1, where $\mathbb{NN}(\cdot)$ and $\mathbb{K}(\cdot)$ represent the parameters of the NN and Kalman modules, respectively. 


\begin{algorithm}[!t]
\footnotesize
\caption{Recursive training of NN for AHS.}\label{alg:streaming}
\begin{algorithmic}
  \Procedure{Streaming}{$(\mathbf{Y, R}) \rightarrow \hat{\mathbf{S}}$}  \Comment{$\mathbf{R}$: reference signal}

    \State Randomly initialize or load pre-trained NN model: $\mathbb{NN}(\cdot)$ 
    \State Randomly generate: $\Delta t$, $G$
    \While{$m\leq M$} \Comment{$M$ is the total number of frames}
      \State $\hat{\mathbf{S}}_m \gets \mathbb{NN}(\mathbf{Y}_m, \mathbf{R}_m)$  \Comment{Output at frame $m$}
      \State $\mathbf{X}_{m} \gets \text{Delayed}\_\hat{\mathbf{S}}_m\cdot G$  
      \State $\mathbf{Y}_{m+1} \gets \mathbf{S}_{m+1} + \mathbf{X}_{m}\cdot \mathbf{H} $ \Comment{Update mic. signal}
      	\If{Hybrid method}  \Comment{Update ref. signal}
	 \State  $\hat{\mathbf{D}}_{m+1} \gets \mathbb{K}(\mathbf{Y}_{m+1}, \mathbf{X}_m$)  
	 \Comment{$ \mathbb{K}(\cdot)$: Kalman}
	  \State  $\mathbf{R}_{m+1} \gets \mathbf{E}_{m+1}=\mathbf{Y}_{m+1}-\hat{\mathbf{D}}_{m+1}$
	\Else{~(NN only method)} 
	\State  $\mathbf{R}_{m+1} \gets \mathbf{X}_m$ 
	 \EndIf
	\State  $\hat{\mathbf{S}} \gets \hat{\mathbf{S}}_m$  \Comment{Save processed frame to final output}
    \EndWhile\label{euclidendwhile}
     \State $Loss \gets (\mathbf{S}, \hat{\mathbf{S}})  $ \Comment{Get loss }
    \State$\mathbb{NN}(\cdot) \gets Loss  $  \Comment{Update DNN parameters}
  \EndProcedure
  \end{algorithmic}

\end{algorithm}

\subsection{Model training}

\subsubsection{NN Only Method}


The NN-only method takes frequency-domain microphone signal $\mathbf{Y}_m$ and reference signal $\mathbf{R}_m$ as input to get an estimate of the target signal $\hat{\mathbf{S}}_m$, as described in Algorithm 1:
\begin{flalign}
\textstyle \hat{\mathbf{S}}_m =   \mathbb{NN}(\mathbf{Y}_m, \mathbf{R}_m)
\end{flalign}
where $m$ denotes the frame index, and the loudspeaker signal obtained in the previous frame $\mathbf{X}_{m-1}$ is used as the reference signal.

\subsubsection{Hybrid Method}
Diagram of the hybrid method is shown in Fig.~\ref{fig:OnlineAHS}(b). It combines NN with traditional Kalman filter, where the Kalman module addresses howling suppression by modeling the acoustic path with an adaptive filter and then subtracting the corresponding estimated playback signal $\hat{\mathbf{D}}_{m} $ from microphone recording to get an error signal $\mathbf{E}_m$:
\begin{flalign}
\textstyle \hat{\mathbf{D}}_m &=   \mathbb{K}(\mathbf{Y}_m, \mathbf{X}_{m-1}) \\\nonumber
\textstyle \mathbf{E}_m &=   \mathbf{Y}_m -  \hat{\mathbf{D}}_m
\end{flalign}
The output of Kalman filter is then used as the reference signal for training the NN module in the hybrid method to get an estimate of the target speech:
\begin{flalign}
\textstyle \hat{\mathbf{S}}_m =   \mathbb{NN}(\mathbf{Y}_m, \mathbf{E}_m)
\end{flalign}
The proposed hybrid method can be view as a recursive training adaptation of HybridAHS \cite{zhang2023hybridAHS}. Unlike HybridAHS, which uses pre-processed signals from the Kalman filter during offline training, our method integrates the NN module and Kalman filter within the closed-loop system for frame-by-frame processing. This approach capitalizes on the strengths of both modules while effectively addressing the mismatch problem in HybridAHS.

\subsubsection{Network structure and loss function}

We employ LSTM for complex ratio mask estimation \cite{williamson2015complex} in the proposed NN module. 
It's worth noting that our primary focus in this study is not to introduce new network architectures; instead, our proposed method offers flexibility in choosing the network structure.
The LSTM network utilized in our implementation consists of two hidden layers, each with 300 units, resulting in 1.54 million trainable parameters. The input for model training is a concatenation of $[|\mathbf{Y}|, |\mathbf{R}|, \mathbf{Y}_r, \mathbf{Y}_i]$, with the training target being set as $[\mathbf{S}_r, \mathbf{S}_i]$. Here $|*|$, $*_r$ and $*_i$ represent the magnitude, real, and imaginary spectrograms of the corresponding frequency-domain signals. We use a frame size and frame shift of 8 ms and 4 ms, respectively. All models are trained for 60 epochs with a batch size of 128.

We use mean absolute error (MAE) of real and imaginary spectrograms at the utterance level as loss function for model training: 
\begin{flalign}
Loss =\text{MAE} (\hat{\mathbf{S}}_r, \mathbf{S}_r)+\text{MAE} (\hat{\mathbf{S}}_i, \mathbf{S}_i)
\end{flalign}

\subsection{Convergence issue in recursive training}
Introducing the recursive training of NN for AHS poses challenges, particularly the difficulty in achieving convergence. The inherent recursive nature of howling generation can lead to signal accumulation and energy explosion, surpassing the maximum allowable value in Python and triggering ``not a number'' (NAN) warnings, hindering gradient calculations and model updates. This issue is especially prominent during batch training, where the convergence failure of one utterance affects the entire batch's loss value. 
To address this challenge, we propose two strategies: howling detection and initialization using pre-trained models.

\subsubsection{Howling detection (HD)}



An effective strategy is to integrate howling detection into the training process. During recursive training, we continuously monitor the microphone signal for the presence of howling, identified by the amplitude of microphone signal consistently exceeding a threshold for 100 consecutive samples. Upon detection, further processing of the current utterance is halted, and only the already processed portion is used for loss calculation.
Excluding the howling signal from further processing and loss calculation prevents potential NAN issue and minimizes its impact on the convergence of the NN module.



\subsubsection{Initialization using pre-trained models}
The other strategy we proposed for enhancing trainability and expediting training involves utilizing a pre-trained offline model to initialize the NN parameters.
Normally, the NN module's parameters are initialized randomly, which may not guarantee adequate howling suppression and can lead to severe howling and NAN warnings during the initial training phases.
Despite the inevitable mismatches in the recursive inference scenarios, the offline pre-trained model still demonstrates superior howling suppression compared to randomly initialized NN modules.
Adopting pre-trained offline models for NN parameter initialization addresses the NAN issue and ensures the convergence of model training.
This approach can be seen as a form of recursive fine-tuning of the offline model. In our implementation, we employ the pre-trained HybridAHS \cite{zhang2023hybridAHS} for NN initialization.

\begin{table*}[!t]
\centering
\caption{Streaming evaluation of different methods for howling suppression. }
\label{table:offline}
\resizebox{0.99 \textwidth}{!}{
\begin{tabular}{c|cccc|cccc} \specialrule{1.5 pt}{1 pt}{1 pt}
Models            & \multicolumn{4}{c|}{SDR (dB)} & \multicolumn{4}{c}{PESQ} \\ \hline
G                    & 1.5   & 2   & 2.5  & 3  & 1.5   & 2   & 2.5 & 3   \\ \hline
no AHS             &  -30.51 $\pm$ 7.23   &  -31.86 $\pm$ 5.66 & -33.10 $\pm$ 3.96   &   -33.21 $\pm$ 3.94  & -- & --    &  --     &  --      \\ \hline
Kalman filter \cite{enzner2006frequency}            &  -5.11 $\pm$13.20 &  -10.33 $\pm$ 14.84  & -14.88 $\pm$ 15.14  &  -18.25  $\pm$ 14.77   &  1.94 $\pm$ 0.72  & 1.65 $\pm$ 0.73    & 1.44 $\pm$ 0.70&  1.30  $\pm$ 0.64 \\
DeepMFC \cite{zheng2022deep}       &    -0.09 $\pm$ 6.50      &   -2.78 $\pm$ 9.44 &  -5.59 $\pm$ 11.40 &     -7.69 $\pm$ 12.26   &  2.11 $\pm$ 0.51 &  1.88 $\pm$ 0.59  & 1.70 $\pm$ 0.62   & 1.56  $\pm$ 0.59 \\ 
HybridAHS \cite{zhang2023hybridAHS}    & 2.96 $\pm$ 3.04   &   1.25 $\pm$ 5.79  &  -1.45 $\pm$ 9.60  &     -3.49   $\pm$10.90 & 2.57 $\pm$ 0.47 &  2.33 $\pm$ 0.53   &2.22 $\pm$ 0.59  &    1.95  $\pm$ 0.62   \\ \hline
Proposed NN     & 3.70 $\pm$ 1.70 & 2.85 $\pm$ 1.45&  2.34 $\pm$ 1.26 &  1.99   $\pm$1.05   & 2.50 $\pm$ 0.43  &    2.28 $\pm$ 0.39 &  2.12 $\pm$ 0.36  & 2.00 $\pm$ 0.34\\  
Proposed Hybrid (RM)      &  2.95 $\pm$ 2.02  &   1.92  $\pm$ 1.70 & 1.28 $\pm$ 1.47  &  0.84 $\pm$ 1.30 & 2.56 $\pm$ 0.40 &  2.35 $\pm$ 0.36   & 2.21 $\pm$ 0.34  & 2.11 $\pm$ 0.32  \\  
Proposed Hybrid      & $\mathbf{3.87}$ $\pm$ $\mathbf{1.68}$   &   $\mathbf{3.04}$  $\pm$ $\mathbf{1.34}$ & $\mathbf{2.49}$ $\pm$ $\mathbf{1.11}$ &  $\mathbf{2.11}$ $\pm$ $\mathbf{0.98}$  &  $\mathbf{2.60}$ $\pm$ $\mathbf{0.41}$ &  $\mathbf{2.40}$  $\pm$ $\mathbf{0.38}$   & $\mathbf{2.25}$ $\pm$ $\mathbf{0.36}$    &  $\mathbf{2.13}$ $\pm$ $\mathbf{0.34}$ \\  \specialrule{1.5 pt}{1 pt}{1 pt}
\end{tabular}
}
\end{table*}


\section{Experimental Setup} 
\subsection{Data Preparation} 

The experiments are conducted using the AISHELL-2 dataset \cite{du2018aishell}. We generate 10,000 pairs of room impulse responses (RIRs) using the image method \cite{allen1979image} with random room characteristics and reverberation times (RT60) randomly selected within the range of 0 to 0.6 seconds. Each RIR pair include RIRs for the near-end speaker and loudspeaker positions. 
Model training follows the steps outlined in Algorithm 1. During training, for each training sample, we randomly select a speech sample and a pair of RIRs for generating the target speech and playback signal. The system delay $\Delta t$ is randomly generated within the range of 0.15 to 0.25 seconds, while the amplification gain $G$ is selected randomly within the range of 1 to 3.

The training, validation, and testing set we used includes 38,000, 1000, and 200 utterances, respectively. The testing data uses different utterances and RIRs compared to the training and validation data.

\subsection{Evaluation Metrics} 

Two metrics presented as $mean$ $\pm$ $standard~deviation$ are used to evaluate AHS performance: signal-to-distortion ratio (SDR) \cite{le2019sdr} and perceptual evaluation of speech quality (PESQ) \cite{rix2001perceptual}. Given PESQ's insensitivity to scale, we emphasize SDR results to demonstrate the effectiveness of suppressing howling, while relying on PESQ to assess speech quality.

\section{Experimental Results}

\subsection{Convergence explorations}

\begin{figure}[!t]
\centering
     \includegraphics[width=0.99\columnwidth]{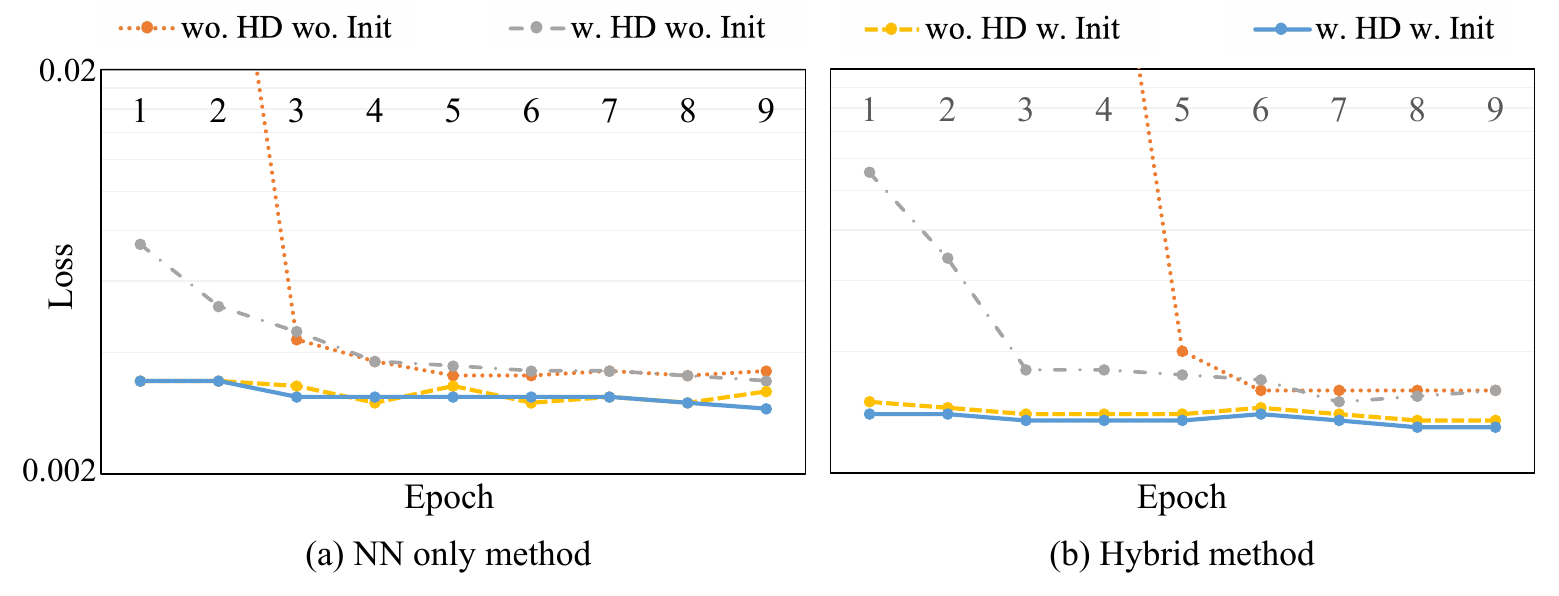}
      \caption{Convergence exploration of the proposed method.}
      \label{fig:loss}
\end{figure}

We first evaluate the effectiveness of our proposed strategies in achieving convergence during recursive training. The loss values across the first few epochs are depicted in Fig.~\ref{fig:loss}. Without the adoption of any strategies, convergence is not assured due to the aforementioned NAN issue. Incorporating the HD strategy successfully circumvents this problem, ensuring the model's trainability. Additionally, initializing the NN module using a pre-trained model (HybridAHS) leads to substantial improvements in the model's convergence.

%
%
%
%
%
%
%
%

\subsection{Howling suppression performance}

\begin{figure}[!t]
\centering
     \includegraphics[width=1\columnwidth]{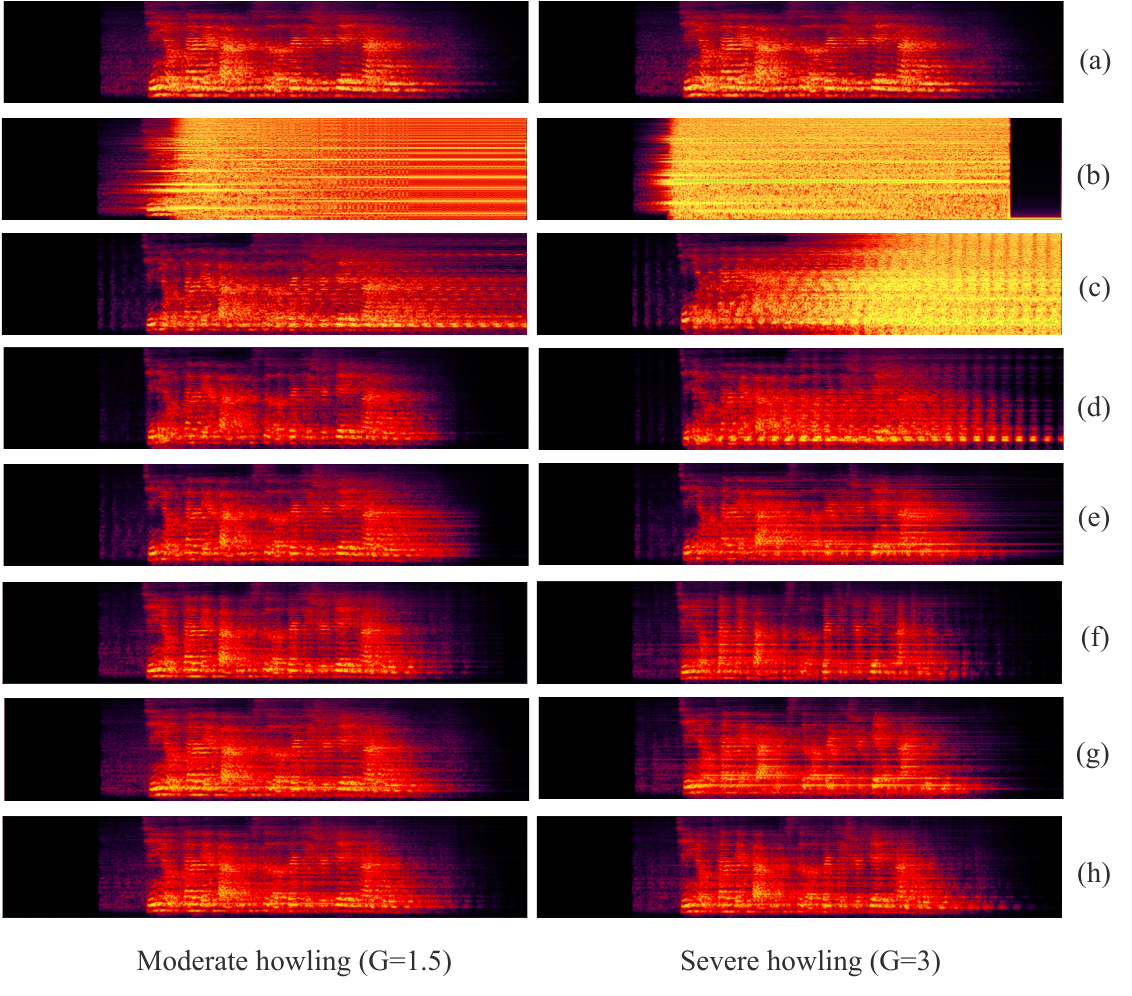}
      \caption{Spectrograms of: (a) target signal, (b)
no AHS, (c) Kalman filter \cite{enzner2006frequency}, (d) DeepMFC \cite{zheng2022deep}, (e) HybridAHS \cite{zhang2023hybridAHS}, (f) Proposed NN, (g) Proposed Hybrid (RM), and (h) Proposed Hybrid. (Demos are available at \href{https://yixuanz.github.io/AHS_2023_1}{https://yixuanz.github.io/AHS\_2023\_1}).}
      \label{fig:spectro}
\end{figure}


We compare the proposed method with Kalman filter and NN-based AHS method across different amplification gain ($G$) levels and present the results in Table~\ref{table:offline}. The spectrograms of a test utterance assessed under moderate and severe howling scenarios are provided in Fig.~\ref{fig:spectro}. All NN-based methods shared the same network architecture and dataset for a fair comparison. 
It is worth mentioning that the proposed recursive training requires longer training times but maintains the same inference times as baseline methods using offline training.


Without any howling suppression (``no AHS''), the microphone signal exhibits average SDR values below 30 dB, indicating significant howling dominance and negligible discernible speech. Therefore, calculating PESQ values in this context becomes redundant.

Utilizing the Kalman filter achieves notable howling suppression compared to the ``no AHS'' case. Moving to NN-based AHS methods (DeepMFC and HybridAHS) results in substantial enhancements in howling suppression. Our proposed approach consistently outperforms the baseline methods in terms of SDR, particularly at higher $G$ levels. In terms of PESQ results, our methods demonstrate comparability to HybridAHS at lower $G$ levels and outperform the baselines at higher $G$ values. Notably, our hybrid method surpasses the NN-only approach in performance.

We also implement the proposed Hybrid method using ratio mask estimation (RM) for comparison. Incorporating complex-domain estimation effectively mitigates playback leakage in the enhanced signal, resulting in improved AHS performance, as also observed in Fig.~\ref{fig:spectro}(g) and (h).

Additionally, our proposed recursive training approach shows lower standard deviations compared to offline-trained baseline methods, indicating enhanced stability and consistent howling suppression.


\section{Conclusion}
In this study, we have introduced a recursive training approach for NN-based AHS, utilizing howling detection and pre-trained model initialization to improve model trainability. Our approach is implemented in both NN-only and Hybrid configurations. The proposed method successfully addresses the mismatch issue observed in previous NN-based AHS techniques, resulting in superior howling suppression while preserving speech quality. This study makes notable contributions to the field of howling suppression. Future directions include exploring alternative recursive training strategies and extending our approach to multi-channel AHS.

\clearpage

\bibliographystyle{IEEEbib}
\bibliography{onlineAHS}

\end{document}